\documentclass[prl,aps,showpacs,twocolumn,floatfix,superscriptaddress]{revtex4}

\newcommand{\beq}{\begin{equation}}
\newcommand{\eeq}{\end{equation}}
\newcommand{\beqs}{\begin{eqnarray}}
\newcommand{\eeqs}{\end{eqnarray}}

\newcommand{\Tr}{{\rm Tr}}

\renewcommand{\L}{{\cal L}}

\def\hbar{\hspace{0pt}\raisebox{1pt}{$-$} \hspace{-7pt} h}

\begin{document}
\title{Quark Mass Ratios and Mixing Angles from $SU(3)$ Family Gauge Symmetry}

\author{Thomas Appelquist \thanks{email: thomas.appelquist@yale.edu}}
\author{Yang Bai \thanks{email: yang.bai@yale.edu}}
\affiliation{Department of Physics, Sloane Laboratory, Yale
University, New Haven, CT 06520}
\author{Maurizio Piai \thanks{email:piai@u.washington.edu }}
\affiliation{Department of Physics, University of Washington,
Seattle, WA 98195}


\begin{abstract}

We explore a framework for the computation of quark mass ratios and CKM
mixing angles based on an $SU(3)$ family gauge symmetry. The four ratios
$m_d/m_b$, $m_s/m_b$, $m_u/m_t$, and $m_c/m_t$ can be fit at one-loop in the
family gauge interaction. The same is true of the quark mixing angles
$\theta_{12}$ and $\theta_{23}$, although the result for $\theta_{13}$ is
too small. The CP violating phase is naturally ${\cal O}(1)$.

\end{abstract}


\maketitle

{\em \bf{Introduction}---}%
We examine the radiative generation of quark mass-ratios and mixing angles
when the standard model (SM) with three families of quarks and leptons is
enlarged to include an SU(3) family gauge interaction \cite{SU(3)refs, Ross,
Others}. We take the family symmetry to be broken at some scale $F$, large
enough to suppress flavor-changing neutral currents. With electroweak
breaking described by a Higgs-doublet field, some mechanism, such as the
inclusion of  supersymmetry, must be invoked to stabilize the Higgs mass. We
do not address this problem here. We assume that any additional new physics,
such as that associated with grand unification, appears only above the
breaking scale $F$ \cite{unification}. We employ an effective field theory
(EFT) to describe flavor physics below this scale. With the family gauge
coupling weak enough, the family gauge bosons are part of the EFT, and their
effects can be computed perturbatively.

{\em \bf{The Model}---}%
To compute the ratios $m_d / m_b$, $m_s / m_b$, $m_u /m_t$, $m_c / m_t$, and
the CKM mixing angles radiatively in the family gauge interaction, these
quantities must vanish in its absence. To this end, we introduce two global
symmetries, $SU(3)_1 \times SU(3)_2$, with the standard model fermions and a
set of partners (the "visible" sector) transforming according to $SU(3)_1$,
and additional fields of a "hidden" sector transforming according to
$SU(3)_2$. The $SU(3)$ family gauge interaction arises from gauging the
diagonal subgroup of $SU(3)_1 \times SU(3)_2$. We also introduce an
additional, discrete $Z_3$ symmetry.

The breaking pattern in the visible sector preserves two $Z_2$ subgroups of
$SU(3)_1$, and can be shown to arise naturally from a broad class of
potentials. This pattern leads to the vanishing of the mixing angles and
quark-mass ratios in the absence of the gauge interaction. The breaking
pattern in the hidden sector also preserves two $Z_2$ subgroups -- of
$SU(3)_2$. The unbroken discrete symmetries are misaligned, such that no
$Z_2$ symmetry remains when the sectors are gauge coupled. The gauge
coupling therefore leads to non-vanishing values for the mass ratios and
mixing angles. Phases of order unity also arise naturally in the breaking in
each sector.

The fields of our model, together with their transformation properties under
the $SU(3)_1 \times SU(3)_2 \times Z_3$ symmetries and the SM symmetries,
are shown in Table~\ref{tab:fields}.
         \begin{table}
\begin{tabular}{||c|c|c|c|c|c|c||}
          \hline
           & $SU(3)_1$ & $SU(3)_2$ & $Z_3$ & $SU(3)_c$ & $SU(2)_L$ & $U(1)_Y$
           \\ \hline \hline
         $q$  & 3 & 1 & $1''$ & 3 & 2 & $\frac{1}{6}$ \\ \hline
         $u^c$  & 3  &1  &$1'$  & $\bar{3}$ & 1 & $-\frac{2}{3}$ \\ \hline
         $d^c$  &3  & 1 & 1 & $\bar{3}$ & 1 & $\frac{1}{3}$ \\ \hline
         $\chi$  &  3 & 1 & $1'$ & 3 & 1 & $\frac{2}{3}$ \\ \hline
         $\chi^c$  & 3 & 1 & 1 & $\bar{3}$ & 1 & $-\frac{2}{3}$ \\ \hline  \hline
         $h$  & 1 &  1 & 1 & 1 & 2 & $-\frac{1}{2}$ \\  \hline
         $S$    & $\bar{6}$ & 1 & $1'$ & 1 & 1 & 0 \\  \hline
         $\Sigma$  & $\bar{6}$ & 1 & $1''$ & 1 &  1& 0 \\  \hline  \hline
         $H$  & 1 & $\bar{6}$ & 1 & 1 & 1 & 0 \\ \hline
          \hline
\end{tabular}
\caption{Field content and symmetries of the model. The $Z_{3}$ labels refer
to the three cube roots of unity. All fermions are LH chiral fields.}
\label{tab:fields}
\end{table}
In addition to the SM fermion fields, there are two fermionic fields, $\chi$
and $\chi^c$, with the SM quantum numbers of the up-type quarks. They
differentiate the up- and down-sectors, and play an important role in
up-type mass generation. Each fermion transforms as a $\bf{3}$ under
$SU(3)_1$, meaning that this symmetry must be broken to generate fermion
mass. In addition to the usual Higgs scalar $h$, two scalar (SM-singlet)
multiplets, $S$ and $\Sigma$, both $\bf{\bar{6}}$'s (symmetric tensors)
under $SU(3)_1$, couple to the fermions. Finally, we describe the hidden
sector by one (SM-singlet) scalar multiplet, $H$, a $\bf{\bar{6}}$ under
$SU(3)_2$.

The EFT for physics below the family breaking scale is comprised of the
fermions, the Higgs boson, the SM gauge bosons, the family gauge bosons, and
the components of the $S$ and $\Sigma$ fields that survive as
pseudo-Goldstone bosons (PGB's). The $SU(3)$ family gauge interaction is
universal with respect to the up-type and down-type fermions. It is, so far,
anomalous, requiring the existence of additional, heavy fermions to remove
the anomalies. When integrated out, they  generate an appropriate
Wess-Zumino-Witten (WZW) term in the EFT \cite{WZW} which must be included
in the analysis. It will not affect the fermion mass ratios and mixing
angles to leading order, and we will not discuss it further here.

Since we are interested in the generation of the Yukawa couplings of the
Standard Model, we focus on operators that are bilinear in the fermion
fields. The operators allowed by the symmetries (including the $Z_3$), and
involving only a single power of $S$ or $\Sigma$, are given by \beqs
\label{Lag-tree} \L_Y =
y_d\frac{qhSd^c}{F}+y_1\frac{q\tilde{h}S\chi^c}{F}+y_2\chi
Su^c+y_3\chi\Sigma \chi^c +{\rm h.c.}
         \eeqs
In the EFT below the cutoff $M_F \equiv 4\pi F$, nonlinear constraints on
$S$ and $\Sigma$ insure that they describe only GB and PGB degrees of
freedom. Each of the $y_i$ couplings is a dimensionless parameter determined
by physics above $M_F$. Each except for $y_1$ will be small compared to the
family gauge coupling $g$, which will be ${\cal O}(1)$, that is, $\alpha/\pi
\equiv g^2/4\pi^2 = {\cal O}(1/40)$. This will justify using the $y_i$
couplings at only lowest order, with quantum corrections arising from the
family gauge interactions. Operators bilinear in the fermions fields but
with higher powers of $S$ and $\Sigma$ are also allowed by the symmetries.
We argue that they naturally produce small effects due to the smallness of
the $y_i$ couplings, after describing the role of the above operators.

{\em \bf{Symmetry Breaking}---}%
We assume that the visible-sector physics above the family breaking scale is
such as to give the following VEV's for $S$ and $\Sigma$:
         \beqs
         \label{tree-vev}
\langle S\rangle\,=\,F\left(
              \begin{array}{ccc}
               0 & 0 & 0 \\
                0 & 0 & 0 \\
                0 & 0 & s \\
              \end{array}
            \right)
            \quad
\langle\Sigma\rangle\,=\,F\left(
              \begin{array}{ccc}
                0 & 0 & 0 \\
                0 & \sigma & 0 \\
                0 & 0 & 0 \\
              \end{array}
            \right).
         \eeqs
         Here $s$ and $\sigma$ are complex numbers of roughly unit magnitude.
They preserve a $Z_2^2$ subgroup of $SU(3)_1$, generated, for example, by
\beqs \label{discrete}
P_1^{(1)}&=&diag\{1,-1,-1\} \nonumber \\
P_1^{(2)}&=&diag\{-1,1,-1\}, \eeqs and also break the discrete $Z_3$
symmetry \cite{Domainwalls}.

To justify this pattern of symmetry breaking, it is helpful to analyze
potential terms with the high energy theory decribed by a linearized theory
of the scalars $S$ and $\Sigma$. The potential couples the $S$ and $\Sigma$
fields, containing terms even in these fields as well as terms such as
$\Tr[(S \times S) S]$ and $\Tr[(\Sigma \times \Sigma) \Sigma]$, where $S
\times S$ denotes the $6$ in the product of the two $\bar{6}$'s. These terms
preserve $SU(3)_1 \times Z_3$, but explicitly break $U(1)_{S}$ and
$U(1)_{\Sigma}$ associated with $S$ and $\Sigma$.

The pattern Eq.~(\ref{tree-vev}), a special case of the general class of
$Z_2^2$-preserving vacua, which can have arbitrary diagonal entries, emerges
for a wide class of potentials \cite{vacuum alignment}. For $S$, an example
is the potential $V = (\Tr SS^* -F^2)^2 + \lambda \Tr [(S \times S)(S \times
S)^*] + \kappa F \Tr[(S \times S)S] + h.c. +...$, where $\lambda$ and
$\kappa$ are dimensionless parameters. Taking $\langle S \rangle$ diagonal
by convention, this potential leads to the above form providing only that
$\lambda > 0$ and that the cubic term is not too large. Similar terms for
the $\Sigma$ field, together with dimension-4 coupling terms such as $\Tr[(S
\Sigma^*)(S^* \Sigma)]$ (with positive definite coefficient), lead to a
diagonal form for $\langle \Sigma \rangle$, also with a single entry, and
prefer to anti--align it with $\langle S \rangle$. Taking the non-zero
entries to be $33$ and $22$, we have the form Eq.~(\ref{tree-vev}). This
pattern can be thought of as preserving two, distinct $SU(2)$ symmetries for
the $S$ and $\Sigma$ sectors separately, with the coupling of the sectors
preferring to mis-align them, leaving no unbroken continuous subgroup of the
common $SU(3)_1$. Eight Goldstone bosons (GB's) are formed. For a similar
construction see ~\cite{Ross}.

The hidden-sector potential is also taken to include terms which preserve
$SU(3)_2$ but explicitly break $U(1)_{H}$. In the absence of the gauge
coupling, the sextet $H$ is  assumed to develop a VEV $\langle H \rangle$ of
${\cal O}(F)$, but leaving no continuous subgroup of the $SU(3)_2$. This,
too, is natural depending on the parameters of the potential. With only one
sextet in this sector, since its VEV can be diagonalized, a $Z_2^2$ subgroup
of the $SU(3)_2$ necessarily remains unbroken.

When the hidden and visible sectors are coupled through the family gauge
interaction, with $\langle S \rangle$ and $\langle \Sigma \rangle$ diagonal
as above, $\langle H \rangle$ takes a general form:
         \beq
         \label{ab}
\langle H\rangle\,=\,F\left(
              \begin{array}{ccc}
              b_1^2 & b_2 & b_3 \\
               b_2 & a_1 & a_3 \\
               b_3 & a_3 & a_2 \\
              \end{array}
            \right),
         \eeq
where $a_i$ and $b_i$ are dimensionless complex numbers. We assume here that
the potential terms gnerated by the family gauge interactions prefer this
pattern with its non-vanishing off-diagonal elements \cite{APY1}. It means
that no discrete ($Z_2^2$) subgroup of the $SU(3)$ family gauge group
remains.

We also assume the existence of a moderate hierarchy in $\langle H \rangle$,
where each $|a_i| = {\cal O}(a)= {\cal O}(1)$ and each $|b_i| = {\cal O}(b)<
{\cal O}(1)$. This form, with one diagonal element quadratic in $b$, emerges
naturally from certain potential terms, for example those that constrain
$|\det \langle H \rangle |^2$ to be ${\cal O}(b^4)$. (These terms by
themselves respect a global $U(1)_{H}$, an approximate symmetry of the full
potential if the explicit breaking terms are somewhat smaller.) The
orientation of $\langle H \rangle$ relative to $\langle S \rangle$ and
$\langle \Sigma \rangle$, with the suppression factor $b$ appearing in
$H_{ij}$ for either index equal to $1$, also emerges naturally from a broad
class of such potential terms. In the limit $b \rightarrow 0$, $\langle H
\rangle$ preserves an (approximate) $U(1)$, a product of $U(1)_{H}$ and a
$U(1)$ subgroup of $SU(3)_2$. This limit also preserves a single, {\bf
exact} $Z_2$ subgroup of the $SU(3)$ family gauge group. Thus the breaking
is sequential, first preserving this $Z_2$ symmetry, and then breaking it.

The hidden sector produces $8$ GB's. They combine with the $8$ GB's of the
visible sector to produce $8$ exact GB's which are eaten by the family gauge
bosons. The other 8 combinations are PGB's, which acquire a (small) mass
through the explicit symmetry breaking of the family gauge interaction. We
discuss the effect of the PGB's later.

{\em \bf{Quark Mass Matrices at Tree Level}---}%
After electroweak breaking, the down-type quark mass matrix is
         \beqs
         M_d= y_d v \frac{\langle S \rangle}{F} = y_d\,v\left(
              \begin{array}{ccc}
               0 & 0 & 0 \\
                0 & 0 & 0 \\
                0 & 0 & s \\
              \end{array}
            \right),
         \eeqs
         where $v \approx 250 GeV$ is the VEV of Higgs doublet $h$. At
this
          level, only the
         $b$ quark develops a mass, of the right order for $y_d \approx
         10^{-2}$. This pattern relies on
         the $Z_3$ symmetry, forbidding the direct, dimension-5 coupling of
$\Sigma$ to the down sector.

         The up-type quark mass matrix is $6\times 6$:
         \beqs
         \label{seesaw}
(u\;\chi) \tilde{M}_u \left(
          \begin{array}{c} u^c \\ \chi^c \\ \end{array} \right)
         =(u\;\chi) \left(
                    \begin{array}{cc}
                       0 & y_1\,v\frac{\langle S\rangle}{F} \\
                      y_2\langle S\rangle &y_3\langle\Sigma\rangle \\
                    \end{array}
                  \right)
         \left(  \begin{array}{c}u^c \\\chi^c \\
          \end{array}\right).
         \eeqs
The squares of the eigenvalues of this (non-symmetric) matrix can be read
off from the diagonal matrix $\tilde{M}_{u} \tilde{M}_u^{\dagger}$. There
are three non-vanishing eigenvalues of order $y_1^2 v^2$, $y_2^2 F^2$, and
$y_3^2 F^2$. The first corresponds to the $t$-quark. That is, its
left-handed component is the $SU(2)_L$-doublet $t$-field. The latter two are
very large providing only that $ v/F \ll y_2, y_3$, and correspond
completely to $SU(2)_L$ singlets.

Thus, in the absence of the family gauge interaction, there are no family
mixings, no masses for the $u$, $d$, $c$, and $s$ quarks, and no mass for
one additional up-type, $SU(2)_L$-singlet fermion.

{\em \bf{SU(3) Family Gauge Interactions}---}%
   The corrections to the quark mass matrices depend directly on
the mass matrix for the family gauge bosons. It arises from the kinetic
terms of the three scalar sextets: \beqs \L_K &=&
\frac{1}{2}\Tr[(D_{\mu}S)(D^{\mu}S)^{*}]\,+\,\frac{1}{2}\Tr[(D_{\mu}\Sigma)(D^{\mu}\Sigma)^{*}]
\nonumber \\ &&+\, \frac{1}{2}\Tr[(D_{\mu}H)(D^{\mu}H)^{*}],\eeqs with
$D_{\mu}G=\partial_{\mu} G+igA_{\mu a}t_aG+igA_{\mu a}Gt_a^*$, where
$t_a,\,a=1,\cdots,8$ are the generators of the $SU(3)$ family gauge symmetry
and $G$ represents the three scalars $S$, $\Sigma$ and $H$. In terms of the
scalar VEV's, the gauge-boson mass operator is then
         \beqs
\L_M = \frac{1}{2}A_a(M^2_{S ab}+M^2_{\Sigma ab}+M^2_{H ab})A_b,
         \eeqs
where, for example,
         \beqs
M^2_{S ab}&=&g^2\Tr[t_a\langle S\rangle t^*_b\langle
S\rangle^*+t_at_b\langle S\rangle \langle S\rangle^*]
          +(a\leftrightarrow b). \nonumber\\
\eeqs

The entries of $M^2_{S ab}+M^2_{\Sigma ab}+M^2_{H ab}$ are of order $g^2
F^2$, but, due to the form of $\langle H \rangle$, some off-diagonal terms
have additional suppression factors expressed by powers of $b$.

{\em \bf{Radiative Corrections to the Quark Mass Matrices}---}%
Perturbation theory is valid providing that $g^{2}/4\pi^2 \ll 1$. We will
require $g \approx 1$ phenomenologically. The loop corrections may be viewed
as corrections to $\langle S \rangle$ and $\langle \Sigma \rangle$ in
Eq.~(\ref{tree-vev}). At one loop, we find \beq \label{radiative} \delta
\langle S \rangle_{ij}\,=\,-\frac{\alpha}{\pi}sF
\log(\frac{M^2_c}{M_F^2})(t_a)^3_i(t_b)^3_j O_{ac}O_{bc},
      \eeq
where $i,j = 1,2,3$ are the family indices and $a,b,c = 1,\cdots,8$ label
the 8 gauge bosons. $M_F$ is the cutoff scale and the $M_c^2$ are the mass
eigenvalues of the family gauge bosons. The matrix $O$ is the orthogonal
transformation diagonalizing the gauge boson mass matrix. The small
parameter $b$ appears in this matrix. The $M_F$ dependence survives in only
the $33$ element, giving a cut-off dependent renormalization of $s$. A
similar expression obtains for $\delta\langle \Sigma \rangle_{ij}$, with the
index $3$ replaced by $2$. To derive these expressions, it is easiest to
work in a renormalizable gauge such as Landau gauge.

In addition to the above contributions to $\delta \langle S \rangle_{ij}$
and $\delta\langle \Sigma \rangle_{ij}$, there are contributions from
fermion wave-function renormalization (kinetic-energy mixing). They lead to
corrections of the same general form, and we don't exhibit them explcitly.

The corrected forms of the $\langle S \rangle$ and $\langle \Sigma \rangle$
matrices thus include ${\cal O}(\alpha/\pi)$ entries replacing the $0$'s in
Eq.~(\ref{tree-vev}). The presence of factors $b$ and $b^2$ in the first row
and column of $\langle H \rangle$ leads to a similar presence in the
corrected $\langle S \rangle$:
         \beqs \label{Sprime}
\langle S\rangle'=\langle S\rangle + \delta \langle S \rangle = F\left(
              \begin{array}{ccc}
               {\cal
O}(\frac{\alpha}{\pi}b^2) & {\cal O}(\frac{\alpha}{\pi}b) &{\cal
O}(\frac{\alpha}{\pi}b) \\
               {\cal
O}(\frac{\alpha}{\pi}b) &{\cal O}(\frac{\alpha}{\pi}) & {\cal
O}(\frac{\alpha}{\pi}) \\
               {\cal
O}(\frac{\alpha}{\pi}b) &{\cal
O}(\frac{\alpha}{\pi}) & s \\
              \end{array}
            \right).\eeqs
The complex coefficient in each entry depends on the values of the
parameters in $\langle H \rangle$. Having explicitly exhibited $b$, all
these coefficients are ${\cal O}(1)$, and can be expressed as functions of
the parameters $b_i$ and $a_j$.

Similarly, the form of the corrected $\langle \Sigma \rangle$ matrix is
\beqs \langle \Sigma \rangle'=\langle \Sigma \rangle + \delta \langle \Sigma
\rangle = F\left(
              \begin{array}{ccc}
             {\cal
O}(\frac{\alpha}{\pi}b^2)    & {\cal O}(\frac{\alpha}{\pi}b) & {\cal
O}(\frac{\alpha}{\pi}b) \\
                {\cal
O}(\frac{\alpha}{\pi}b) & \sigma & {\cal
O}(\frac{\alpha}{\pi}) \\
               {\cal
O}(\frac{\alpha}{\pi}b) & {\cal O}(\frac{\alpha}{\pi}) & {\cal
O}(\frac{\alpha}{\pi}) \\
              \end{array}
            \right).
         \eeqs
Here again, each entry in the first row and column carries suppression
factor of ${\cal O}(b)$.

The mass matrix for the down-type quarks is $ M_d = y_{d} v \langle S
\rangle' /F$. Diagonalizing this matrix, we obtain the mass ratios for
down-type quarks. To leading order in $\alpha/\pi$, we find \beqs
\label{md/mb} \frac{m_d}{m_b} &\approx&
\frac{\alpha}{\pi}b^2 \\
\label{ms/mb}
        \frac{m_s}{m_b} &\approx& \frac{\alpha}{\pi} , \eeqs with the
        $b$-quark mass given by $m_b \approx y_{d}v$. We have dropped
corrections
        of
        ${\cal O}(b^2)$ in each expression. Each
        includes a coefficient of ${\cal O}(1)$ arising from physics above $M_F$.

The up-type masses are obtained from Eq.~(\ref{seesaw}) using the corrected
forms $\langle S \rangle'$ and $\langle \Sigma \rangle'$. We determine the
eigenvalues and mixing angles from the $6 \times 6$ symmetric matrix
         \beq \label{MMdaggertilde}
\tilde{M}_u \tilde{M}^\dagger_u= \left(
\begin{array}{cc}
y^2_1v^2\frac{\langle S\rangle' \langle S\rangle'^\dagger}{F^2} &
y_1y_3v\frac{\langle S\rangle' \langle \Sigma\rangle'^\dagger}{F} \\
         y_1y_3v\frac{\langle \Sigma \rangle'\langle S \rangle'^\dagger}{F} &
         y^2_2\langle S\rangle'\langle
         S\rangle'^\dagger+y_3^2\langle\Sigma\rangle'
\langle\Sigma\rangle'^\dagger \\
         \end{array}
\right).
         \eeq
The three heavy eigenvalues are ${\cal O}(y_{2}^{2}F^2)$, ${\cal
O}(y_{3}^{2}F^2)$, and ${\cal O}((\alpha/\pi)^2F^{2}b^{2}(y_3^2+y_2^2))$.
Each is well beyond experimental reach for the range of parameters
considered here. We integrate them out, leading to the following $3 \times
3$ matrix for the up-type quarks:
         \beqs \label{MMdagger}
         M_{u}M^\dagger_u&=&y^2_1v^2\frac{\langle
S\rangle'}{F}\Big(I+\frac{{\langle\Sigma\rangle'}^{\dagger} (\langle S
\rangle' \langle S \rangle'^{\dagger})^{-1}\langle\Sigma\rangle'}{z^2}
\Big)^{-1}\frac{{\langle S\rangle'}^\dagger}{F}, \nonumber \\
         \eeqs
valid for $v/F \ll y_2\alpha/\pi~,~ y_3 \alpha/\pi $. Here, $z\equiv
y_2/y_3$.

Diagonalizing this matrix gives the up-type quark masses and mixing angles.
To lowest non-vanising order in $\alpha/\pi$, they take simple algebraic
forms. In the limit $(\alpha^2/\pi^2)b^2 \ll z^2 \ll b^2 \ll 1$, appropriate
for our numerical fits, we find \beqs \label{mu/mt} \frac{m_u}{m_t}&\approx&
\frac{\alpha^2}{\pi^2}bz  \\
\label{mc/mt} \frac{m_c}{m_t}&\approx& \frac{\alpha}{\pi} \frac{z}{b}. \eeqs
Each expression includes a coefficient of ${\cal O}(1)$ arising from physics
above $M_F$. The $t$-quark mass is given by $ m_t \approx y_{1}vb$. Recall
that for $\alpha = 0$, the mass-eigenvalue of order $y_1 v$ corresponded to
the $SU(2)_L$-doublet $t$-field. As $\alpha$ is increased to $(\alpha/\pi)b
\gg v/F$, this eigenvalue grows to be ${\cal
O}((\alpha/\pi)^2F^{2}b^{2}(y_3^2+y_2^2))$, but its $t$-component decreases
to nearly zero. Meanwhile, one of the zero mass-eigenvalues at $\alpha = 0$
grows to ${\cal O}(y_{1}vb)$, with its $t$-component growing from zero to
nearly $100 \%$.

The (small) CKM mixing angles emerge as differences between the
diagonalization angles for $M_{d} M_{d}^{\dagger}$ and $M_{u}
M_{u}^{\dagger}$. Using conventional definitions \cite{PDG}, we find, again
to lowest non-vanishing order in $\alpha/\pi$ and in the limit
$(\alpha^2/\pi^2)b^2 \ll z^2 \ll b^2 \ll 1$,
         \beqs
         \label{theta23}
         \theta_{23}&\approx& \frac{\alpha}{\pi}
         \\
         \label{theta13}
         \theta_{13}&\approx&\frac{\alpha^2}{\pi^2}b
         \\
        \label{theta12}
         \theta_{12}&\approx& b.      \eeqs
The ${\cal O}(1)$ phases appearing throughout the mass matrices naturally
generate an ${\cal O}(1)$ CKM phase.

Before turning to the phenomenology, it is worth noting that although the
expressions for $m_u/m_t$ and $\theta_{13}$ are ${\cal O}(\alpha^2/\pi^2)$,
they are one-loop results. The form of $m_u/m_t$ is due simply to the
product of $\alpha$ factors appearing in the "seesaw" expression
 Eq.~(\ref{MMdagger}) for $M_uM_u^{\dagger}$. This can be seen directly by
computing $\det M_uM_u^{\dagger}$, or by computing $\det \tilde{M}_u
\tilde{M}^\dagger_{u}$ Eq.~(\ref{MMdaggertilde}). The latter quantity can be
shown to be proportional to $(\alpha/\pi)^{8} b^{8} z^{6} v^{6} F^6$ in the
limit $(\alpha^2/\pi^2)b^2 \ll z^2 \ll b^2 \ll 1$. The values of the three
large eigenvalues of this matrix are as reported above, with one
proportional to $\alpha^2$. But only three of its six eigenvalues vanish as
$g \rightarrow 0$, and one ($m_c^2$) is proportional to $\alpha^2$. Thus
$m_u^2$ must be proportional to $\alpha^4$. If higher-order corrections are
added to the expressions for $\langle S\rangle'$ and $\langle \Sigma
\rangle'$, this will not change the leading-order value of $\det\tilde{M}_u
\tilde{M}^\dagger_{u}$. Thus the expression Eq.~(\ref{mu/mt}) for $m_u/m_t$
will not be affected by two-loop contributions.

We also find $\theta_{13} =  {\cal O}(\alpha^2/\pi^2)$ Eq.~(\ref{theta13}),
again a one-loop result.  We compute $\theta_{13}$ by first noting that the
up-sector mixing angles are determined by the second term in the large
brackets in Eq.~(\ref{MMdagger}), that is, from the approximate expression
$M_u \sim \langle S\rangle' \langle \Sigma \rangle'^{-1} \langle S\rangle'$.
Using this expression in a basis in which $\langle S\rangle'$ is diagonal,
we have $\theta_{13}= M_{u13}/M_{u33}- M_{u12}M_{u23}/M_{u22}M_{u33}$. Each
term is separately ${\cal O}(\alpha/\pi)$, but they cancel exactly, leaving
$\theta_{13} = {\cal O}(\alpha^2/\pi^2)$. The cancellation is a consequence
of only the fact that the ${\cal O}(\alpha/\pi)$ corrected form of $\langle
S \rangle'$ Eq.~(\ref{Sprime}) is up-down universal, relying in no way on
the details of the family gauge interaction. Just as with $m_u/m_t$, the
result for $\theta_{13}$ is not modified by including higher-order
corrections in $\langle S\rangle'$ and $\langle \Sigma \rangle'$.

{\em \bf{Phenomenology}---}%
Each of the $7$ approximate expressions
(\ref{md/mb},\ref{ms/mb},\ref{mu/mt},\ref{mc/mt},\ref{theta23},\ref{theta13},\ref{theta12})
depends on one or more of the $3$ small parameters $\alpha/\pi$, $b$, and
$z$. Each except $\theta_{12}$ vanishes as $\alpha \rightarrow 0$. The
coefficients of order unity in each expression depend on the ${\cal O}(1)$
parameters in $\langle H \rangle$, $\langle S \rangle$, and $\langle \Sigma
\rangle$.

We compare the above expressions to the measured values of the mixing angles
and the quark mass ratios~\cite{PDG}. These depend of course on the scale at
which they are defined. The Yukawa interactions derived here should be
regarded as defined at the scale $M_F \gg v$, to be evolved to lower scales
through SM interactions. We disregard these renormalization group effects
here, and simply compare our expressions with the quark masses and CKM
angles defined at $M_Z$. The quark masses in GeV units are $m_t(M_Z) = 176
\pm 5$, $m_b(M_Z) = 2.95 \pm 0.15$, $m_c(M_Z) = 0.65 \pm 0.12$, $m_s(M_Z) =
0.062 \pm 0.015$, $m_u(M_Z) = 0.0017 \pm 0.0005$, $m_d(M_Z) = 0.0032 \pm
0.0009$. The mixing angles measured in tree-level processes, and as defined
in Ref. \cite{PDG}, are: $\sin \theta_{12} = 0.2243 \pm 0.0016$, $\sin
\theta_{23} = 0.0413 \pm 0.0015$, $\sin \theta_{13} = 0.0037 \pm 0.0005$.

The $6$ approximate expressions
(\ref{md/mb},\ref{ms/mb},\ref{mu/mt},\ref{mc/mt},\ref{theta23},\ref{theta12})
(excluding the expression for $\theta_{13}$), can be fit to the data, up to
coefficients of order unity, by the choices
        \beqs \label{alpha} \frac{\alpha}{\pi} &\approx&
0.04
         \\
         \label{b}
         b &\approx& 0.2
         \label{y}
         \\
         z &\approx& 0.04 .
         \eeqs
While the $6$ expressions
(\ref{md/mb},\ref{ms/mb},\ref{mu/mt},\ref{mc/mt},\ref{theta23},\ref{theta12})
used in the fit are accurate for these values, we have also performed a
numerical study of the model, using the complete expressions for each of the
mass ratios and mixing angles, and found that a good agreement with the
experimental values can be obtained with the above choices. Finally, the $t$
and $b$ masses are fit with $y_1 \approx 4$ and $y_d \approx 10^{-2}$.

The mixing angle $\theta_{13}$, however, is not well fit by the expression
Eq.~(\ref{theta13}). It is too small by nearly an order of magnitude
assuming that the coefficient is ${\cal O}(1)$. Since this is a one-loop
result in the gauge interaction, and since none of the couplings of our
effective theory is larger than the (weak) gauge coupling, there appears to
be no ingredient in the theory that could produce a large enough value for
$\theta_{13}$. Additional ingredients may be necessary to generate a
$\theta_{13}$ at the measured (${\cal O}(4\times 10^{-3})$) level. This
problem is currently under study.

{\em \bf{Discussion}---}%
There are additional corrections to the quark mass matrices from the
emission and re-absorption of the $8$ PGB's of the EFT , whose
masses can be estimated to be ${\cal O}((g^2 /4 \pi)F)$. Their
coupling to the light fermions is suppressed by $v/F$, but they
couple to the up-type heavy fermions with strength $y_2$ or $y_3$.
Their corrections to the up-type mass matrices can thus be as large
as ${\cal O}(y_3^2 / 4 \pi^2)$. These will be smaller than the
family gauge corrections providing only that $y_3 \ll g \approx 1$,
as we have already assumed.

In addition to the operators bilinear in the fermion fields that we have
analyzed so far (Eq.~(\ref{Lag-tree})), there are more such operators, with
higher powers of $S$ and $\Sigma$. The nonlinear constraints of the EFT mean
that they are not dimensionally suppressed, and they can contribute directly
to the quark mass matrices. Among the operators with two powers of the
scalar fields, because of the structure of $\langle S \rangle$ and $\langle
\Sigma \rangle$, only one contributes to the quark masses to zeroth order in
$\alpha$. It is $q\tilde{h}(S\times \Sigma)^*u^c / F^2$. Since $\langle
S\times \Sigma \rangle = diag\{s \sigma F^2,0,0\}$ to zeroth order, it gives
a direct contribution to the mass of up quark. Its presence reflects the
fact that the diagonal zero's in $\langle S \rangle$ and $\langle \Sigma
\rangle$ Eq.~(\ref{tree-vev}) are not stable against quantum corrections.

The natural size of this operator can be argued, however, to be small. It
can be estimated from quantum loops based on the interactions of
Eq.~(\ref{Lag-tree}). We first note that such quantum loops do not
destabilize the assumed smallness of the $y_i$ themselves. They are
protected by chiral symmetries in the limit $y_i \rightarrow 0$. For the
operator $q\tilde{h}(S\times \Sigma)^*u^c / F^2$, its coefficient can be
estimated to be ${\cal O}(y_1 y_2 y_3 / 4 \pi^2)$. For the range of $y_i$
values employed here, this gives a contribution to $m_u$ comfortably smaller
than the family-gauge contribution. Similar analyses can be applied to the
full tower of operators bilinear in the fermion fields.

It is important to extend the framework suggested here to the charged
leptons and neutrinos. A neutrino seesaw mechanism could emerge using the
large scale $F$, providing it is large enough. It is also possible that the
present framework can be combined with grand unification at still higher
scales \cite{unification}.

{\em \bf{Conclusion}---}%
We conclude that the general pattern of up-type quark masses, down-type
quark masses, and CKM mixing angles, with the exception of $\theta_{13}$,
can be understood as arising radiatively from a relatively weak $SU(3)$
family gauge interaction, with a sequential breaking of this symmetry --
first preserving a discrete $Z_2$ subgroup of the gauged $SU(3)$ and then
breaking it. The up-type quark mass ratios are generated via mixing with
heavy fermions after the radiative corrections are included. The detailed
predictions depend on the two additional small parameters $b$ (in the scalar
potential of the hidden sector) and $z$ (a ratio of two Yukawa couplings of
the up-type fermions), as well as various ${\cal O}(1)$ parameters, all of
which are determined by physics above the family breaking scale. The values
of $m_b$ and $m_t$ are determined by the parameters $y_d$ and $y_1$, which
also arise from physics above the family breaking scale. The framework
leads, however, to a value for $\theta_{13}$ that is smaller than the
measured value by nearly an order of magnitude, indicating the need for
additional ingredients.

\begin{acknowledgments}
This work was partially supported by Department of Energy grants
DE-FG02-92ER-4074 (T.A. and Y.B.) and DE-FG02-96ER40956 (M.P.). We
thank Richard Easther, Walter Goldberger, Zhenyu Han, Ann Nelson,
Pierre Ramond, Robert Shrock, and Witold Skiba for helpful
discussions. One of us (M.P.) thanks Yale University for its
hospitality during the completion of this work.
\end{acknowledgments}



\begin{thebibliography}{99}

\bibitem{SU(3)refs} On $SU(3)$ family symmetry:
M.~Bowick and P.~ Ramond, Phys. Lett.\ B {\bf 103}, 338 (1981);
D.~R.~T.~Jones, G.~L.~Kane and J.~P.~Leveille,
  Nucl.\ Phys.\ B {\bf 198}, 45 (1982);
Z.~G.~Berezhiani,
         Phys.\ Lett.\ B {\bf 150}, 177 (1985);
 Z.~G.~Berezhiani and M.~Y.~Khlopov,
  Sov.\ J.\ Nucl.\ Phys.\  {\bf 51}, 739 (1990)
  [Yad.\ Fiz.\  {\bf 51}, 1157 (1990)];
Z.~Berezhiani and A.~Rossi,
       Nucl.\ Phys.\ B {\bf 594}, 113 (2001)
       [arXiv:hep-ph/0003084];
      A.~Masiero, M.~Piai, A.~Romanino and L.~Silvestrini,
       Phys.\ Rev.\ D {\bf 64}, 075005 (2001)
       [arXiv:hep-ph/0104101];
       G.~G.~Ross, L.~Velasco-Sevilla and O.~Vives,
       Nucl.\ Phys.\ B {\bf 692}, 50 (2004)
       [arXiv:hep-ph/0401064].
A.~Hernandez Galeana and J.~H.~Montes de Oca Yemha,
  Rev.\ Mex.\ Fis.\  {\bf 50}, 522 (2004)
  [arXiv:hep-ph/0406315].


       \bibitem{Ross}    S.~F.~King and G.~G.~Ross,
       Phys.\ Lett.\ B {\bf 520}, 243 (2001)
       [arXiv:hep-ph/0108112];
       Phys.\ Lett.\ B {\bf 574}, 239 (2003)
       [arXiv:hep-ph/0307190].

\bibitem{Others} On other non-abelian family symmetries:
  M.~Dine, R.~G.~Leigh and A.~Kagan,
  Phys.\ Rev.\ D {\bf 48}, 4269 (1993)
  [arXiv:hep-ph/9304299];
  D.~B.~Kaplan and M.~Schmaltz,
  Phys.\ Rev.\ D {\bf 49}, 3741 (1994)
  [arXiv:hep-ph/9311281];
  L.~J.~Hall and H.~Murayama,
  Phys.\ Rev.\ Lett.\  {\bf 75}, 3985 (1995)
  [arXiv:hep-ph/9508296];
A.~Pomarol and D.~Tommasini,
        Nucl.\ Phys.\ B {\bf 466}, 3 (1996) [arXiv:hep-ph/9507462];
K.~S.~Babu and R.~N.~Mohapatra,
       Phys.\ Rev.\ Lett.\  {\bf 83}, 2522 (1999)
       [arXiv:hep-ph/9906271];
  H.~Fritzsch and Z.~z.~Xing,
  Prog.\ Part.\ Nucl.\ Phys.\  {\bf 45}, 1 (2000)
  [arXiv:hep-ph/9912358].


\bibitem{unification}
         S.~M.~Barr,
         Phys.\ Rev.\ D {\bf 37}, 204 (1988);
      R.~Barbieri, L.~J.~Hall, S.~Raby and A.~Romanino,
       Nucl.\ Phys.\ B {\bf 493}, 3 (1997)
       [arXiv:hep-ph/9610449];


\bibitem{WZW}
      J.~Wess and B.~Zumino,
       Phys.\ Lett.\ B {\bf 37}, 95 (1971);
      E.~Witten,
       Nucl.\ Phys.\ B {\bf 223}, 422 (1983);


\bibitem{Domainwalls} Spontaneously broken discrete symmetries can lead to
problems with cosmological domain walls. Various mechanisms have been
discussed to deal with this. See, for example, G.~Lazarides and
Q.~Shafi, 
  Phys.\ Rev.\ D {\bf 27}, 995 (1983)
 and C.~Panagiotakopoulos and
K.~Tamvakis,
  Phys.\ Lett.\ B {\bf 446}, 224 (1999)
  [arXiv:hep-ph/9809475].


\bibitem{vacuum alignment}
L.~F.~Li,
         Phys.\ Rev.\ D {\bf 9}, 1723 (1974).



\bibitem{APY1}
T.~Appelquist, Y.~Bai and M.~Piai,
  Phys.\ Rev.\ D {\bf 72}, 036005 (2005)
  [arXiv:hep-ph/0506137].



\bibitem{PDG}
S.~Eidelman {\it et al.}  [Particle Data Group],
         Phys.\ Lett.\ B {\bf 592}, 1 (2004).


\end{thebibliography}
\end{document}